\documentstyle[12pt]{article}

\def\noi{\noindent}

\def\nqq{\hspace{-2em}}

\def\barr{\left(\begin{array}}
\def\earr{\end{array}\right)}
\def\beq#1{\begin{equation}\label{#1}}
\def\eeq{\end{equation}}
\def\ber#1{\begin{eqnarray}\label{#1} \nqq}%   left alignment
\def\eer{\end{eqnarray}}
\def\eern{\nonumber \end{eqnarray}}%   nonumber
\def\nn{\nonumber\\ \nqq}%   left alignment
\def\mm{\\ \nqq}

\catcode`\@=11 \@addtoreset{equation}{section}\catcode`\@=12

\newcommand{\R}{\mbox{\bf R}}
\newcommand{\Z}{\mbox{\bf Z}}
\newcommand{\N}{\mbox{\bf N}}

\newcommand{\diag}{\mathop{\rm diag}\nolimits}
\newcommand{\sign}{\mathop{\rm sign}\nolimits}

\newcommand{\eps}{\varepsilon}
\newcommand{\tri}{\triangle}

\newcommand{\e}[1]{\mathop{\rm e}\nolimits^{#1}}

\newcommand{\p}{\partial}

\newcommand{\fnm}{\footnotemark}
\newcommand{\fnt}{\footnotetext}

\begin{document}

\begin{center}
\large\bf
%\\
HYPERBOLIC KAC-MOODY ALGEBRA
>FROM INTERSECTING P-BRANES\footnote{This work was
supported by the Russian Ministry of Science and
Technology and by RFBR grant 98-02-16414.} \\[15pt]

\normalsize\bf V.D. Ivashchuk\fnm[2]\fnt[2]{ivas@rgs.phys.msu.su},
 S.-W. Kim
\fnm[3]\fnt[3]{sungwon@mm.ewha.ac.kr} \\[10pt]
and
V.N. Melnikov
\fnm[4]\fnt[4]{rgs@com2com.ru} \\[10pt]

\it Center for Gravitation and Fundamental Metrology,
VNIIMS, 3/1 M. Ulyanovoy Str.,
Moscow 117313, Russia${}^{2,4}$

\it Department of Science Education
and Basic Science Research Institute, Ewha Womans University, Seoul 120-750,
Korea${}^3$
\end{center}

\vspace{15pt}

\small\noi

\begin{abstract}

A subclass of
recently discovered class of solutions in multidimensional gravity
with intersecting $p$-branes related to Lie algebras and governed by
a set of harmonic functions is considered. This subclass
in case of three Euclidean $p$-branes (one electric and two magnetic)
contains a cosmological-type solution
(in $11$-dimensional model with two $4$-forms)
related to hyperbolic Kac-Moody algebra ${\cal F}_3$ (of rank $3$).
This solution describes the non-Kasner power-law inflation.
\end{abstract}

\vspace{10cm}

\hspace*{0.950cm} PACS number(s):\ 04.50.+h,\ 98.80.Hw,\ 04.60.Kz

%\pagebreak
\normalsize

%%%%%%%%%%%%%%%%%%%%%%%%%%%%%%%%%%%%%%%%%%%%%%%%%%%%%%%%%%%%%%%%
\section{Introduction}
%%%%%%%%%%%%%%%%%%%%%%%%%%%%%%%%%%%%%%%%%%%%%%%%%%%%%%%%%%%%%%%%

Here we consider recently discovered class of solutions
with intersecting $p$-branes \cite{IMBl}. These solutions are governed
by a set of harmonic functions. The number of harmonic functions
in general is less then the number of $p$-branes (as it takes
place in orthogonal case \cite{S}-\cite{AIR}).
The solutions  correspond to a block-orthogonal
set of $p$-brane vectors $U^s$ (see (\ref{4.17}) below) and
may be considered as a Majumdar-Papapetrou type extension of
for the extremal limit  of  ``block-orthogonal'' black holes recently
found in \cite{Br} (for ``orthogonal'' black holes see
\cite{CT,OS,AIV,O,BIM,IMJ} and references therein.)

For 1-block case the solution is governed by one harmonic
function and for a special configuration  may be related to some
simple finite dimensional Lie algebra or infinite dimensional
hyperbolic Kac-Moody (KM) algebra  \cite{Kac,FS}. The affine KM algebras
do not appear among the solutions from \cite{IMBl}.

Let us consider the simplest example of $D = 11$ supergravity
\cite{CJS} (corresponding to M-theory \cite{M-th1}).
It is known \cite{Str,To,Ts1} that
the orthogonal (or $A_1 + A_1$) intersection rules for the M-theory
read
\ber{1.1}
3\cap3 =1, \quad 3\cap6=2, \quad 6 \cap 6 =4
\eer
(here we are counting dimensions of world-sheets and their
intersections). For the  simplest $A_2 = sl(3)$  Lie algebra   the
intersection rules are modified as follows
\cite{IMBl}
\ber{1.2}
3\cap3 =0, \quad 3\cap6=1, \quad 6 \cap 6 =3.
\eer
The  rules (\ref{1.2})
are obtained from (\ref{1.1}) by a shift of one unit.
(For $3\cap3 =0$ the ``truncated'' theory
or without Chern-Simons term   should be considered).
These modified rules may be writen
for a wide class of  models and Lie algebras (finite or hyperbolic) and are
defined by a Dynkin diagrams \cite{IMBl,IMJ}.

Hyperbolic algebras  appeared in  different areas of
mathematical physics, e.g. in ordinary gravity \cite{FF}
(${\cal F}_3$ hyperbolic algebra),  supergravity:
\cite{J,Miz} ($E_{10}$ hyperbolic algebra),
\cite{Nic1} (${\cal F}_3$ hyperbolic algebra),
strings etc
(see also \cite{Nik} and references therein).
In \cite{Nic1} it was shown that the  chiral
reduction of a simple ($N=1$) supergravity  from four dimensions
to one dimension gives rise to the hyperbolic algebra of rank
3 (namely ${\cal F}_3$).

In \cite{IMBl}  we considered some examples of hyperbolic
intersection rules  for the hyperbolic KM algebras of
rank 2. These examples were suggested for so-called
$B_D$ models with $D \geq 14$ \cite{IMJ}, containing
$D-11$ scalar fields with negative kinetic terms.
($B_{11}$ is the truncated bosonic sector of $D =11$ supergravity.
$B_{12}$ is the 12-dimensional  model  \cite{KKLP}
corresponding to
the low energy limit of F-theory \cite{F-th}.)

Here  an example of cosmological solution
in $11$-dimensional model  with three p-branes (two magnetic
and one electric)  that have intersection rules
corresponding to the hyperbolic KM algebra ${\cal F}_3$
is constructed.

%%%%%%%%%%%%%%%%%%%%%%%%%%%%%%%%%%%%%%%%%%%%%%%%%%%%%%%%%%%%%%%%
\section{The model}
%%%%%%%%%%%%%%%%%%%%%%%%%%%%%%%%%%%%%%%%%%%%%%%%%%%%%%%%%%%%%%%%

We consider a  model governed by
the action \cite{IMC}
\ber{4.1}
S=\int d^Dz\sqrt{|g|}\biggl\{R[g]-h_{\alpha\beta}g^{MN}\p_M\varphi^\alpha
\p_N\varphi^\beta-\sum_{a\in\tri}\frac{\theta_a}{n_a!}
\exp[2\lambda_a(\varphi)](F^a)^2\biggr\}
\eer
where $g=g_{MN}dz^M\otimes dz^N$ is a metric,
$\varphi=(\varphi^\alpha)\in\R^l$ is a vector of scalar fields,
$(h_{\alpha\beta})$ is a symmetric
non-degenerate $l\times l$ matrix $(l\in \N)$,
$\theta_a=\pm1$, $F^a=dA^a$ is a $n_a$-form ($n_a\ge1$), $\lambda_a$ is a
1-form on $\R^l$: $\lambda_a(\varphi)=\lambda_{\alpha a}\varphi^\alpha$,
$a\in\tri$, $\alpha=1,\dots,l$. Here $\tri$ is some finite set.

We consider a manifold
\ber{4.2}
M=M_0\times M_1\times\dots\times M_n,
\eer
with a metric
\ber{4.3}
g=\e{2\gamma(x)}g^0+\sum_{i=1}^n\e{2\phi^i(x)}g^i
\eer
where $g^0=g_{\mu\nu}^0(x)dx^\mu\otimes dx^\nu$ is a metric on the
manifold $M_0$, and $g^i=g_{m_in_i}^i(y_i)dy_i^{m_i}\otimes dy_i^{n_i}$
is a  Ricci-flat  metric on the manifold $M_i$ ($Ric[g^i]= 0$),
$i=1,\dots,n$. Any manifold $M_\nu$ is
oriented and connected and $d_\nu\equiv\dim M_\nu$,
$\nu=0,\dots,n$. Let
\ber{4.5}
\tau_i \equiv\sqrt{|g^i(y_i)|}dy_i^1\wedge\dots\wedge dy_i^{d_i}, \quad
\eps(i)\equiv\sign(\det(g_{m_in_i}^i))=\pm1
\eer
denote the volume $d_i$-form and signature parameter respectively,
$i=1,\dots,n$. Let $\Omega=\Omega_n$ be a set of all subsets of
$\{1,\dots,n\}$, $|\Omega|=2^n$. For any $I=\{i_1,\dots,i_k\}\in\Omega$,
$i_1<\dots<i_k$, we denote
\ber{4.6}
\tau(I)\equiv\tau_{i_1}\wedge\dots\wedge\tau_{i_k},
\quad d(I)\equiv\sum_{i\in I}d_i,
\quad \eps(I) \equiv \prod_{i \in I} \eps(i).
\eer
We also put $\tau(\emptyset)= \eps(\emptyset)=
1$ and $d(\emptyset)=0$.

For fields of forms we consider the following composite electromagnetic
ansatz
\ber{4.7}
F^a=\sum_{I\in\Omega_{a,e}}{\cal F}^{(a,e,I)}+
\sum_{J\in\Omega_{a,m}}{\cal F}^{(a,m,J)}
\eer
where
\ber{4.8}
{\cal F}^{(a,e,I)}=d\Phi^{(a,e,I)}\wedge\tau(I), \mm
\label{4.9}
{\cal F}^{(a,m,J)}=\e{-2\lambda_a(\varphi)}*(d\Phi^{(a,m,J)}
\wedge\tau(J))
\eer
are elementary forms of electric and magnetic types respectively,
$a\in\tri$, $I\in\Omega_{a,e}$, $J\in\Omega_{a,m}$ and
$\Omega_{a,e}\subset\Omega$, $\Omega_{a,m}\subset\Omega$. In (\ref{4.9})
$*=*[g]$ is the Hodge operator on $(M,g)$. For scalar functions we put
\ber{4.10}
\varphi^\alpha=\varphi^\alpha(x), \quad
\Phi^s=\Phi^s(x),
\eer
$s\in S$.

Here and below
\ber{4.11}
S=S_e\sqcup S_m, \quad
S_v=\bigcup_{a\in\tri}\{a\}\times\{v\}\times\Omega_{a,v},
\eer
$v=e,m$.

Due to (\ref{4.8}) and (\ref{4.9})
\ber{4.12}
d(I)=n_a-1, \quad d(J)=D-n_a-1,
\eer
for $I\in\Omega_{a,e}$, $J\in\Omega_{a,m}$.

\subsection{The sigma model}

Let $d_0 \neq 2$ and
\ber{4.13}
\gamma=\gamma_0(\phi) \equiv
\frac1{2-d_0}\sum_{j=1}^nd_j\phi^j,
\eer
i.e. the generalized harmonic gauge is used.

We impose the restriction on sets $\Omega_{a,v}$.
These restrictions guarantee the block-diagonal structure
of a stress-energy tensor (like for the metric) and the existence of
$\sigma$-model representation \cite{IMC}.

We denote $w_1\equiv\{i|i\in\{1,\dots,n\},\quad d_i=1\}$, and
$n_1=|w_1|$ (i.e. $n_1$ is the number of 1-dimensional spaces among
$M_i$, $i=1,\dots,n$).

{\bf Restriction 1.} Let 1a) $n_1\le1$ or 1b) $n_1\ge2$ and for
any $a\in\tri$, $v\in\{e,m\}$, $i,j\in w_1$, $i<j$, there are no
$I,J\in\Omega_{a,v}$ such that $i\in I$, $j\in J$ and $I\setminus\{i\}=
J\setminus\{j\}$.

{\bf Restriction 2} (only for $d_0=1,3$). Let 2a) $n_1=0$ or
2b) $n_1\ge1$ and for any $a\in\tri$, $i\in w_1$ there are no
$I\in\Omega_{a,m}$, $J\in\Omega_{a,e}$ such that $\bar I=\{i\}\sqcup J$
for $d_0 = 1$ and
$J=\{i\}\sqcup \bar I$ for $d_0 = 3$. Here and in what
follows
\ber{4.13a}
\bar I\equiv\{1,\ldots,n\}\setminus I.
\eer

It was proved in \cite{IMC} that equations of motion for the model
(\ref{4.1}) and the Bianchi identities: $d{\cal F}^s=0$, $s\in S_m$, for
fields from (\ref{4.3})--(\ref{4.13}), when Restrictions 1 and 2 are
imposed, are equivalent to equations of motion for the $\sigma$-model
governed by the action
\ber{2.1}
S_\sigma=\int d^{d_0}x\sqrt{|g^0|}\biggl\{R[g^0]-\hat G_{AB}
g^{0\mu\nu}\p_\mu\sigma^A\p_\nu\sigma^B \nn
-\sum_{s\in S}\eps_s\e{-2U_A^s\sigma^A}
g^{0\mu\nu} \p_\mu\Phi^s\p_\nu\Phi^s\biggr\},
\eer
where $(\sigma^A)=(\phi^i,\varphi^\alpha)$, the index set
$S$ from (\ref{4.11}), target space metric
\ber{4.14}
(\hat G_{AB})=\barr{cc}
G_{ij}& 0\\
0& h_{\alpha\beta}
\earr,
\eer
with
\ber{4.15}
G_{ij}= d_i \delta_{ij}+\frac{d_i d_j}{d_0-2},
\eer
vectors
\ber{4.17}
(U_A^s)=(d_i\delta_{iI_s},-\chi_s\lambda_{\alpha a_s}),
\eer
where $s=(a_s,v_s,I_s)$, $\chi_e=+1$, $\chi_m=-1$;
\ber{i}
\delta_{iI}=\sum_{j\in I}\delta_{ij}
\eer
is the indicator of $i$ belonging
to $I$: $\delta_{iI}=1$ for $i\in I$ and $\delta_{iI}=0$ otherwise; and
\ber{4.18}
\eps_s=(-\eps[g])^{(1-\chi_s)/2}\eps(I_s) \theta_{a_s},
\eer
$s\in S$, $\eps[g]\equiv\sign\det(g_{MN})$. More explicitly
(\ref{4.18}) reads $\eps_s=\eps(I_s) \theta_{a_s}$ for
$v_s = e$ and $\eps_s=-\eps[g] \eps(I_s) \theta_{a_s}$, for
$v_s = m$.

%%%%%%%%%%%%%%%%%%%%%%%%%%%%%%%%%%%%%%%%%%%%%%%%%%%%%%%%%%%%%%%%
\subsection{Exact solutions in a block-orthogonal case}
%%%%%%%%%%%%%%%%%%%%%%%%%%%%%%%%%%%%%%%%%%%%%%%%%%%%%%%%%%%%%%%%

Let us  define the scalar product as follows
\ber{2.2}
(U,U')=\hat G^{AB}U_AU'_B,
\eer
for $U,U'\in\R^N$, where $(\hat G^{AB})=(\hat G_{AB})^{-1}$.
The scalar products (\ref{2.2}) for vectors $U^s$  were calculated in
\cite{IMC}
\ber{4.19}
(U^s,U^{s'})=d(I_s\cap I_{s'})+\frac{d(I_s)d(I_{s'})}{2-D}+
\chi_s\chi_{s'}\lambda_{\alpha a_s}\lambda_{\beta a_{s'}} h^{\alpha\beta}
\equiv B^{ss'},
\eer
where $(h^{\alpha\beta})=(h_{\alpha\beta})^{-1}$; $s=(a_s,v_s,I_s)$ and
$s'=(a_{s'},v_{s'},I_{s'})$ belongs to $S$.

Let
\ber{2.3}
S=S_1\sqcup\dots\sqcup S_k,
\eer
$S_i\ne\emptyset$, $i=1,\dots,k$, and
\ber{2.4}
(U^s,U^{s'})=0
\eer
for all $s\in S_i$, $s'\in S_j$, $i\ne j$; $i,j=1,\dots,k$. Relation
(\ref{2.3}) means that the set $S$ is a union of $k$ non-intersecting
(non-empty) subsets $S_1,\dots,S_k$. According to (\ref{2.4}) the set of
vectors $(U^s,s\in S)$ has a block-orthogonal structure with respect to
the scalar product (\ref{2.2}), i.e. it  splits into $k$ mutually
orthogonal blocks $(U^s,s\in S_i)$, $i=1,\dots,k$.

Here we consider exact solutions in the
model (\ref{4.1}), when vectors $(U^s,s\in S)$ obey the block-orthogonal
decomposition (\ref{2.3}), (\ref{2.4}) with scalar products defined in
(\ref{4.19}) \cite{IMBl}. These solutions may be obtained from the
corresponding solutions of the $\sigma$-model \cite{IMBl}, that
are presented in Appendix 1.

The solution reads:
\ber{4.20}
g=U\left\{g^0+\sum_{i=1}^n U_ig^i\right\}, \mm
\label{4.21}
U=\left(\prod_{s\in S}H_s^{2d(I_s)\eps_s\nu_s^2}\right)^{1/(2-D)}, \mm
\label{4.22}
U_i=\prod_{s\in S}H_s^{2\eps_s\nu_s^2\delta_{iI_s}}, \mm
\label{4.23}
\varphi^\alpha=-\sum_{s\in S}\lambda_{a_s}^\alpha\chi_s
\eps_s\nu_s^2\ln H_s, \mm
\label{4.24}
F^a=\sum_{s\in S}{\cal F}^s\delta_{a_s}^a,
\eer
where  ${\rm Ric}[g^0] = {\rm Ric}[g^i] = 0$,
\ber{4.25}
{\cal F}^s=\nu_sdH_s^{-1}\wedge\tau(I_s), \mbox{ for } v_s=e, \mm
\label{4.26}
{\cal F}^s=\nu_s(*_0dH_s)\wedge\tau(\bar I_s), \mbox{ for } v_s=m,
\eer
$H_s$ are harmonic functions on $(M_0,g^0)$ coinciding inside blocks
of matrix $(B^{ss'})$ from (\ref{4.19}) ($H_s=H_{s'}$, $s,s'\in S_j$,
$j=1,\dots,k$) and relations
\ber{4.27}
\sum_{s'\in S} B^{ss'} \eps_{s'}\nu_{s'}^2=-1
\eer
for the matrix $(B^{ss'})$ (\ref{4.19}),
parameters $\eps_s$ (\ref{4.18}) and  $\nu_s$ are imposed, $s\in S$,
$i=1,\dots,n$; $\alpha=1,\dots,l$. Here $\lambda_a^\alpha=
h^{\alpha\beta}\lambda_{\beta a}$, $*_0=*[g^0]$ is the Hodge operator
on $(M_0,g^0)$ and $\bar I$  is defined in (\ref{4.13a}).

In deriving the solutions  the following relations for
contravariant components of $U^s$-vectors were used \cite{IMC}:
\ber{4.29a}
U^{si}=\delta_{iI_s}-\frac{d(I_s)}{D-2}, \quad
U^{s\alpha}=-\chi_s\lambda_{a_s}^\alpha,
\eer
$s=(a_s,v_s,I_s)$.

Thus, we obtained the generalization of the solutions from \cite{IMC} to the
block-orthogonal case (here we eliminate the misprint with sign in eq.
(5.19) in \cite{IMC}).

{\bf Remark 1}. The solution is also valid for $d_0=2$, if Restriction 2
is replaced by Restriction $2^{*}$.

{\bf Restriction 2${}^*$} (for $d_0=2$). For any  $a \in \tri$  there are
no $I\in\Omega_{a,m}$, $J\in\Omega_{a,e}$ such that $\bar I =  J$ and for
$n_1 \geq 2$, $i,j\in w_1$, $i \neq j$, there are no $I\in\Omega_{a,m}$,
$J\in\Omega_{a,e}$ such that $i\in I$, $j\in \bar J$, $I\setminus\{i\}=
\bar J \setminus \{j\}$.

It may be proved using a more general
form of the sigma-model representation (see Remark 2 in \cite{IMC}).

%%%%%%%%%%%%%%%%%%%%%%%%%%%%%%%%%%%%%%%%%%%%%%%%%%%%%%%%%%%%%%%%
\section{Solutions related to Lie algebras and intersection rules}
%%%%%%%%%%%%%%%%%%%%%%%%%%%%%%%%%%%%%%%%%%%%%%%%%%%%%%%%%%%%%%%%

Here we put
\ber{3.1}
(U^s,U^s)\ne0
\eer
for all $s\in S$ and introduce the quasi-Cartan matrix $A=(A^{ss'})$
\ber{3.2}
A^{ss'} \equiv \frac{2(U^s,U^{s'})}{(U^{s'},U^{s'})},
\eer
$s,s'\in S$. From (\ref{2.4}) we get a block-orthogonal structure
of $A$:
\ber{3.3}
A=\barr{ccc}
A_{(1)} \dots 0\\
\vdots \ddots  \vdots\\
0 \dots  A_{(k)}
\earr,
\eer
where $A_{(i)}=(A^{ss'},s,s'\in S_i)$, $i=1,\dots,k$. Here we tacitly
assume that the set $S$ is ordered, $S_1<\dots<S_k$ and the order in $S_i$
is inherited by the order in $S$.

We note that due to (\ref{2.4}) the relation (\ref{2.10}) may
be rewritten as
\ber{2.13}
\sum_{s'\in S_i}(U^s,U^{s'})\eps_{s'}\nu_{s'}^2=-1,
\eer
$s\in S_i$, $i=1,\dots,k$. Hence, parameters $(\nu_s,s\in S_i)$
depend upon vectors $(U^s,s\in S_i)$, $i=1,\dots,k$.

For $\det A_{(i)}\ne0$ relation (\ref{2.13}) may be rewritten in the
equivalent form
\ber{3.4}
\eps_s\nu_s^2(U^s,U^s)=-2\sum_{s'\in S}A_{ss'}^{(i)},
\eer
$s\in S_i$, where $(A_{ss'}^{(i)})=A_{(i)}^{-1}$. Thus, eq. (\ref{2.13})
may be resolved in terms of $\nu_s$ for certain $\eps_s=\pm1$, $s\in S_i$.

In what follows we consider the block-orthogonal decomposition to be
irreducible, i.e. for any $i$ the block $(U^s,s\in S_i)$ does not
split into two mutually orthogonal subblocks. In this case any matrix
$A_{(i)}$ is indecomposable (or irreducible) in the sense that there is
no renumbering of vectors which would bring $A_{(i)}$ to the block
diagonal form $A_i=\diag(A'_{(i)},A''_{(i)})$.

Let $A$ be the generalized Cartan matrix \cite{Kac,FS}. In this case
\ber{3.6}
A^{ss'}\in -\Z_+ \equiv\{0,-1,-2,\dots\}
\eer
for $s \ne s'$ and $A$ generates generalized
symmetrizable Kac-Moody algebra \cite{Kac,FS}.

Now we fix $i\in\{1,\dots,k\}$. From (\ref{3.3}) and (\ref{3.6}) we get
\ber{3.7}
A_{(i)}^{ss'}\in-\Z_+,
\eer
$s,s'\in S_i$, $s \ne s'$.
There are three possibilities for $A_{(i)}$: a) $\det A_{(i)}>0$,
b) $\det A_{(i)}<0$ and c) $\det A_{(i)}=0$. For $\det A_{(i)}\ne0$ the
corresponding Kac-Moody algebra is simple, since $A_{(i)}$ is
indecomposable \cite{FS}.

\subsection{Finite dimensional Lie algebras}

Let $\det A_{(i)}>0$. In this case $A_{(i)}$ is the Cartan matrix of a simple
finite-dimensional Lie algebra and $A_{(i)}^{ss'}\in\{0,-1,-2,-3\}$,
$s\ne s'$. The elements of inverse matrix $A_{(i)}^{-1}$ are positive
(see Ch.7 in \cite{FS}) and hence we get from (\ref{3.4})
\ber{3.8}
\eps_s(U^s,U^s)<0,
\eer
$s\in S_i$.

\subsection{Hyperbolic Kac-Moody algebras}

Let $\det A_{(i)}<0$. Among  irreducible
symmetrizable martrices satisfying (\ref{3.7}) there exists a large
subclass of Cartan matrices, corresponding to infinite-dimensional
simple hyperbolic generalized
Kac-Moody (KM) algebras of ranks $r=2,\dots,10$ \cite{Kac,FS}.

For the  hyperbolic algebras the following
relations are satisfied
\ber{3.16}
\eps_s(U^s,U^s) >0,
\eer
$s\in S_i$. This relation is valid, since
$(A_{(i)}^{-1})_{ss'}\le0$, $s,s' \in S$, for any hyperbolic algebra
\cite{NikP}.

It was shown in \cite{IMBl} that  affine KM algebras
with $\det A_{(i)} = 0$ do not appear in the solutions \cite{IMBl}.

\subsection{Intersection rules}

>From the orthogonality relation (\ref{2.4}) and (\ref{4.19}) we get
\ber{5.1}
d(I_s \cap I_{s'})=\tri(s,s')
\eer
where $s\in S_i$, $s'\in S_j$, $i\ne j$ and
\ber{5.2}
\tri(s,s')\equiv\frac{d(I_s)d(I_{s'})}{D-2}-
\chi_s\chi_{s'}\lambda_{a_s}\cdot\lambda_{a_{s'}}.
\eer
Here $\lambda\cdot\lambda'\equiv h^{\alpha\beta}\lambda_\alpha\lambda'_\beta$.
Let
\ber{5.3}
N(a,b)\equiv\frac{(n_a-1)(n_b-1)}{D-2}-\lambda_a\cdot\lambda_b,
\eer
$a,b\in\tri$. The matrix (\ref{5.3}) is called the fundamental matrix of the
model (\ref{4.1}) \cite{IMJ}. For $s_1,s_2\in S$, $s_1\ne s_2$, the symbol
of orthogonal intersection (\ref{5.2}) may be expressed by means of the
fundamental matrix \cite{IMJ}
\ber{5.4}
\tri(s_1,s_2)=\bar D\bar\chi_{s_1}\bar\chi_{s_2}+
\bar{n}_{a_{s_1}}\chi_{s_1}\bar\chi_{s_2}+
\bar{n}_{a_{s_2}}\chi_{s_2}\bar\chi_{s_1}+
N(a_{s_1},a_{s_2})\chi_{s_1}\chi_{s_2},
\eer
where $\bar D=D-2$, $\bar n_a=n_a-1$, $\bar\chi_s=\frac12(1-\chi_s)$.
More explicitly (\ref{5.4}) reads
\ber{5.4a}
\Delta(s_1,s_2)
=N(a_{s_1},a_{s_2}), \quad v_{s_1}=v_{s_2}=e; \mm
\label{5.4b}
\Delta(s_1,s_2) =
\bar{n}_{a_{s_1}}-N(a_{s_1},a_{s_2}), \quad v_{s_1}=e, \quad v_{s_2}=m; \mm
\label{5.4c}
\Delta(s_1,s_2) =
\bar{D}-\bar{n}_{a_{s_1}}-\bar{n}_{a_{s_2}}+N(a_{s_1},a_{s_2}),
\quad v_{s_1}=v_{s_2}=m.
\eer

This follows from the relations
\ber{5.5}
d(I_s)=\bar D\bar\chi_s+\bar n_{a_s}\chi_s,
\eer
equivalent to (\ref{4.12}). Let
\ber{5.6}
K(a)\equiv n_a-1-N(a,a)=\frac{(n_a-1)(D-n_a-1)}{D-2}+
\lambda_a\cdot\lambda_a,
\eer
$a\in\tri$.

The parameters (\ref{5.6}) play a rather important role in
supergravitational
theories, since they are preserved under Kaluza-Klein reduction \cite{S}
and define the norms of $U^s$ vectors:
\ber{5.7}
(U^s,U^s)=K(a_s),
\eer
$s\in S$.

Here we put $K(a)\ne0$, $a\in\tri$. Then, we obtain the general intersection
rule formulas
\ber{5.8}
d(I_{s_1}\cap I_{s_2})=\tri(s_1,s_2)+\frac12K(a_{s_2})A^{s_1s_2}
\eer
$s_1\ne s_2$, where $(A^{s_1s_2})$ is the quasi-Cartan matrix (\ref{3.2})
(see also (6.32) from \cite{IMJ}).

In most models including $D=11$ supergravity, $D =12$ theory \cite{KKLP},
$D < 11$ supergravities \cite{S},  $K(a)=2$
and (\ref{5.8}) has the following form
\ber{5.13a}
d(I_{s_1}\cap I_{s_2})=\tri(s_1,s_2)+A^{s_1s_2},
\eer
$s_1\ne s_2$, and get $A^{s_1s_2}=A^{s_2s_1}$, i.e. the Cartan matrix is
symmetric.

%%%%%%%%%%%%%%%%%%%%%%%%%%%%%%%%%%%%%%%%%%%%%%%%%%%%%%%%%%%%
\section{Examples}
%%%%%%%%%%%%%%%%%%%%%%%%%%%%%%%%%%%%%%%%%%%%%%%%%%%%%%%%%%%%

\subsection{Hyperbolic algebra of rank three}

Now we consider the example of the solution corresponding to the
hyperbolic KM algebra ${\cal F}_3$ with the Cartan matrix
\ber{6.1}
A=\barr{ccc}
2 &-2 & 0 \\
-2 & 2  & -1 \\
0  & -1 & 2
\earr,
\eer

The hyperbolic algebra  ${\cal F}_3$ corresponding
to (\ref{6.1}), is an infinite dimensional Lie algebra generated by the
(Serre) relations \cite{Kac,FS}
\ber{6.2}
[h_i,h_j] =0, \quad [e_i,f_j] = \delta_{ij} h_j \mm
\label{6.3}
[h_i,e_j] = A_{ij} e_j, \quad [h_i,f_j] = -A_{ij} f_j
\mm
\label{6.4}
({\rm ad} e_i)^{1- A_{ij}}(e_j) = 0 \quad (i \neq j),
\mm
\label{6.5}
({\rm ad} f_i)^{1- A_{ij}}(f_j) = 0 \quad (i \neq j).
\eer

${\cal F}_3$ contains $A_1^{(1)}$
affine Kac-Moody subalgebra  (it corresponds to the Geroch group) and
$A_2$  subalgebra.

The calculation of inverse matrix gives us
\ber{6.6}
A^{-1}= - \barr{ccc}
\frac{3}{2} & 2 &1 \\
2 & 2 & 1 \\
1 & 1 & 0
\earr,
\eer
and, hence,
\ber{6.7}
\sum_{j=1}^{3} A_{ij}^{-1} = - \frac{9}{2}, -5, -2,
\eer
for $i =1,2,3$ respectively.

There exists an example of the solution with the $A$-matrix
(\ref{6.6}) for $11$-dimensional model governed by the
action
\ber{6.8a}
S= \int d^{11}z \sqrt{|g|} \biggl\{R[g] - \frac{1}{4!} (F^4)^2
- \frac{1}{4!} (F^{4*})^2 \biggr\},
\eer
where ${\rm rank } F^{4} = {\rm rank} F^{4*} = 4$.
Here $\Delta = \{ 4, 4* \}$.

We consider a configuration with
two  magnetic $5$-branes  corresponding to the form $F^4$ and
one  electric $2$-brane corresponding to the form  $F^{4*}$.
We denote $S = \{s_1,s_2,s_3 \}$,
$a_{s_1} = a_{s_3} = 4$, $a_{s_2} = 4*$ and
$v_{s_1} = v_{s_3} = m$, $v_{s_2} = e$, where
$d(I_{s_1}) = d(I_{s_3}) = 6$ and $d(I_{s_2}) = 3$.
>From intersection rules (\ref{5.13a}) we obtain
\ber{6.9}
d(I_{s_1} \cap I_{s_2}) = 0, \quad
d(I_{s_2} \cap I_{s_3}) = 1, \quad
d(I_{s_1} \cap I_{s_3}) = 4.
\eer

For the manifold (\ref{4.2}) we put
$n= 5$ and $d_1 =2$, $d_2 =4$, $d_3 = d_4 =1$, $d_5 = 2$.
The corresponding sets for $p$-branes are the following:
$I_{s_1} = \{1,2 \}$, $I_{s_2} = \{4,5 \}$, $I_{s_3} = \{2,3,4 \}$.

The corresponding solution reads
\ber{6.10}
g=H^{-12} \left\{ - dt \otimes dt + H^9 g^1 + H^{13} g^2
+ H^4 g^3  + H^{14} g^4  + H^{10} g^5  \right\}, \mm
\label{6.11a}
F^4= \frac{dH}{dt} \left\{
\nu_{s_1} \tau_3 \wedge \tau_4 \wedge \tau_5 +
\nu_{s_3} \tau_1 \wedge \tau_5 \right\}, \quad
\label{6.11}
F^{4*} = \frac{dH}{dt} \frac{\nu_{s_2}}{H^2} dt \wedge \tau_4 \wedge \tau_5,
\eer
where
\ber{6.12}
\nu_{s_1}^2  = \frac{9}{2}, \quad  \nu_{s_2}^2  = 5,
\quad \nu_{s_3}^2 = 2
\eer
(see relations (\ref{3.4})  and (\ref{6.7})), all metric
$g^i$ are Ricci-flat ($i = 1, \ldots, 5$) with the Euclidean
signature (this agrees with relations (\ref{3.16}) and  (\ref{4.18})),
and
\ber{6.14}
H = ht + h_0 > 0,
\eer
$h, h_0$ are constants. (We remind that here $(U^s,U^s) =2$.)

The metric (\ref{6.10}) may be also rewritten using
the synchronous time variable $t_s$
\ber{6.10c}
g= - dt_s \otimes dt_s + f^{3/5} g^1 + f^{-1/5} g^2
+ f^{8/5} g^3  + f^{-2/5} g^4  + f^{2/5} g^5,
\eer
where $f = 5h t_s = H^{-5} > 0$, $h > 0$, $t_s > 0$.
The metric describes the power-law "inflation" in $D =11$. It is singular
for $t_s \to +0$. It is interesting to note that the powers
in scale-factors  $f^{2 \alpha_i}$
do not satisfy Kasner-like relations \cite{I}:
$\sum_{i=1}^{5} d_i \alpha_i  = \sum_{i=1}^{5} d_i (\alpha_i)^2 = 1$.
For flat $g^i$ the calculation of the Riemann tensor squared gives
us (see \cite{IMA,IMB})
\ber{6.10d}
R_{MNPQ}[g]R^{MNPQ}[g] = A t_s^{-4},
\eer
where $A = 2 \times 1,0714$.

\subsection{$A_3$ Lie algebra}

Here we present for comparison the  solution of  $D =11$ supergravity
corresponding to  $A_3$  Lie algebra with the Cartan matrix
\ber{6.1b}
A=\barr{ccc}
2 &-1 & 0 \\
-1 & 2  & -1 \\
0  & -1 & 2
\earr.
\eer

We remind  that $D = 11$ supergravity is governed by the
action (in the bosonic sector)
\ber{6.8b}
S= \int d^{11}z \sqrt{|g|} \biggl\{R[g] - \frac{1}{4!} (F^4)^2 \biggr\}
+ c \int A^3 \wedge F^4 \wedge F^4
\eer
where $c = {\rm const}$,  $F^4 = d A^3$.

The calculation of inverse matrix gives in this case
\ber{6.7b}
\sum_{j=1}^{3} A_{ij}^{-1} =  \frac{3}{2}, 2, \frac{3}{2},
\eer
for $i =1,2,3$ respectively.

Like in the example mentioned above we consider  three $p$-branes, one
electric and two magnetic, i. e. in this case $S = \{s_1,s_2,s_3 \}$,
$v_{s_1} = v_{s_3} = m$, $v_{s_2} = e$,
$d(I_{s_1}) = d(I_{s_3}) = 6$, $d(I_{s_2}) = 3$.
>From intersection rules (\ref{5.13a}) we obtain
\ber{6.9b}
d(I_{s_1} \cap I_{s_2}) = 1, \quad
d(I_{s_2} \cap I_{s_3}) = 1, \quad
d(I_{s_1} \cap I_{s_3}) = 4.
\eer

For the manifold (\ref{4.2}) we put
$n= 5$ and $d_1 =2$, $d_2 =3$, $d_3 = 1$, $d_4 =2$, $d_5 = 2$.
The corresponding sets for $p$-branes are the following:
$I_{s_1} = \{1,2,3 \}$, $I_{s_2} = \{3,5 \}$, $I_{s_3} = \{2,3,4 \}$.

The corresponding solution reads
\ber{6.10b}
g=H^{16/3} \left\{ d\rho \otimes d\rho + H^{-3} g^1
+ H^{-6} g^2
+ H^{-10} (-dt \otimes dt)  + H^{-3} g^4  + H^{-4} g^5 \right\} \mm
\label{6.11b}
F^4= \frac{dH}{d\rho} \left\{  \nu_{s_1} \tau_4 \wedge \tau_5 +
\frac{\nu_{s_2}}{H^2} d\rho \wedge dt \wedge \tau_5 +
\nu_{s_3} \tau_1 \wedge \tau_5 \right\},
\eer
where
\ber{6.12b}
\nu_{s_1}^2  = \frac{3}{2}, \quad   \nu_{s_2}^2  = 2,
\quad \nu_{s_3}^2 = \frac{3}{2}.
\eer
Here the metrics $g^i$ are Ricci-flat ($i = 1,2,4,5$) with the Euclidean
signature, and
\ber{6.14b}
H = c \rho + c_0 >0,
\eer
$c, c_0$ are constants. So, we obtained the multidimensional
"cosmological" solution with the Euclidean "time"  $\rho$.

The solution (\ref{6.10b})-(\ref{6.14b})
satisfies not only equations of motion for the truncated model
(without the Chern-Simons term), but also  the equations of motion
for the "total" model (\ref{6.8b}), since the only modification
related to "Maxwells" equations
\ber{6.15b}
d*F^4 = {\rm const} \ F^4 \wedge F^4,
\eer
is trivial due to $F^4 \wedge F^4 = 0$ (since $\tau_i \wedge \tau_i =0$).

%%%%%%%%%%%%%%%%%%%%%%%%%%%%%%%%%%%%%%%%%%%%%%%%%%%%%%%%%%%%%%%%
\section{Appendix 1: block-orthogonal solutions in the
$\sigma$-model}
%%%%%%%%%%%%%%%%%%%%%%%%%%%%%%%%%%%%%%%%%%%%%%%%%%%%%%%%%%%%%%%%

Equations of motion corresponding to (\ref{2.1}) have the following
form
\ber{2.5}
R_{\mu\nu}[g^0]=\hat G_{AB}\p_\mu\sigma^A\p_\nu\sigma^B+
\sum_{s\in S}\eps_s
\e{-2U_A^s\sigma^A}\p_\mu\Phi^s\p_\nu\Phi^s, \mm
\label{2.6}
\hat G_{AB}\tri[g^0]\sigma^B +
\sum_{s\in S}\eps_sU_A^s\e{-2U_C^s\sigma^C}
g^{0\mu\nu} \p_\mu\Phi^s\p_\nu\Phi^s =0, \mm
\label{2.7}
\p_\mu\left(\sqrt{|g^0|}g^{0\mu\nu}\e{-2U_A^s\sigma^A}
\p_\nu\Phi^s\right)=0,
\eer
$s\in S$. Here  $\tri[g^0]$ is the Laplace-Beltrami operator
corresponding to $g^0$.

{\bf Proposition \cite{IMBl}.} Let $(M_0,g^0)$ be Ricci-flat
$R_{\mu\nu}[g^0]=0$.
Then the field configuration
\ber{2.9}
g^0, \qquad \sigma^A=\sum_{s\in S}\eps_sU^{sA}\nu_s^2\ln H_s, \qquad
\Phi^s=\frac{\nu_s}{H_s},
\eer
$s\in S$, satisfies the field equations (\ref{2.5})--(\ref{2.7}) with
$V=0$ if (real) numbers $\nu_s$ obey the relations
\ber{2.10}
\sum_{s'\in S}(U^s,U^{s'})\eps_{s'}\nu_{s'}^2=-1
\eer
$s\in S$, functions $H_s >0$ are harmonic, i.e.
\ber{2.11}
\tri[g^0]H_s=0,
\eer
$s\in S$ and $H_s$ are coinciding inside blocks:
\ber{2.12}
H_s=H_{s'}
\eer
for $s,s'\in S_i$, $i=1,\dots,k$.

The Proposition  can be readily verified by a straightforward
substitution of (\ref{2.9})--(\ref{2.12}) into equations of motion
(\ref{2.5})--(\ref{2.7}). In the special (orthogonal) case, when any
block contains only one vector (i.e. all $|S_i|=1$) the Proposition
coincides with Proposition 1 of \cite{IMC}. In general case vectors
inside each
block $S_i$ are not orthogonal. The solution under consideration
depends on $k$ independent harmonic functions. For a given set of
vectors $(U^s,s\in S)$ the maximal number $k$ arises for the irreducible
block-orthogonal decomposition (\ref{2.3}), (\ref{2.4}), when any block
$(U^s,s\in S_i)$ does  not  split into two mutually-orthogonal
subblocks.

%%%%%%%%%%%%%%%%%%%%%%%%%%%%%%%%%%%%%%%%%%%%%%%%%%%%%%%%%%%%%%%
\section{Discussions}
%%%%%%%%%%%%%%%%%%%%%%%%%%%%%%%%%%%%%%%%%%%%%%%%%%%%%%%%%%%%%%%%

Here we obtained the example of the cosmological
solution with three
Euclidean intersecting $p$-branes (one electric and two magnetic)
satisfying intersection  rules  for the hyperbolic Kac-Moody Lie algebra
${\cal F}_3$  (see (\ref{5.13a}) and (\ref{6.1})). The corresponding $A_3$
solution contains three pseudo-Euclidean $p$-branes.
The difference in sign rules  (restriction on $\eps_s$)
for finite and hyperbolic algebras is a consequence
of inequalities for elements of the inverse
Cartan matrix: $A^{-1}_{ij} > 0$ for
simple (or semisimple) finite dimensional Lie algebra and
$A^{-1}_{ij} \leq 0$ (for simple hyperbolic KM algebra).
In this paper the hyperbolic KM algebra ${\cal F}_3$ appeared
only on the simplest level of the Cartan matrix (governing the intersection
rules) but the full
structure of the algebra, including Serre relations (\ref{6.4})
and (\ref{6.5}), was not used. We may suppose that at the
second step a more deep penetrating into a "structure"
of infinite dimensional hyperbolic algebras will be achieved
when  general cosmological solutions related to
hyperbolic Toda-lattices will be considered.

\begin{center}
{\bf Acknowledgments}
\end{center}

This work was supported in part
by the Russian Ministry for
Science and Technology, Russian Foundation for Basic Research,
Korea Research Foundation made in the program year of 1977
and KOSEF No. 95-0702-04-01-3.
The authors  are grateful to M.A. Grebeniuk for usefull discussions
and V.V. Nikulin for useful information.
V.D.I and V.N.M also thanks  the members of
Department of Science Education and Basic Science Research Institute
of Ewha Womans University (Seoul, Korea) for the kind hospitality
during their visits in February of 1998. The authors
are grateful to organizers and participants of the Second Winter School
on Branes, Fields and Mathematical Physics (APCTP, Seoul), where
the results of this paper were reported.

\end{document}